\documentclass[twocolumn]{aastex631}

\usepackage{amsmath}
\usepackage{mathtools}
\usepackage{comment}

\begin{document}

\title{The double white dwarf merger progenitors of SDSS J2211+1136 and ZTF J1901+1458}



\author[0000-0002-5438-3460]{M. F. Sousa}
\affiliation{Divis\~ao de Astrof\'{\i}sica, Instituto Nacional de Pesquisas Espaciais, Avenida dos Astronautas 1758, 12227--010, S\~ao Jos\'e dos Campos, SP, Brazil}
\affiliation{ICRANet-Ferrara, Dip. di Fisica e Scienze della Terra, Universit\`a degli Studi di Ferrara, Via Saragat 1, I--44122 Ferrara, Italy}

\author[0000-0001-9386-1042]{J. G. Coelho}
\affiliation{Divis\~ao de Astrof\'{\i}sica, Instituto Nacional de Pesquisas Espaciais, Avenida dos Astronautas 1758, 12227--010, S\~ao Jos\'e dos Campos, SP, Brazil}
\affiliation{N\'ucleo de Astrof\'{\i}sica e Cosmologia (Cosmo-Ufes) \& Departamento de F\'isica, Universidade Federal do Esp\'irito Santo, 29075--910, Vit\'oria, ES, Brazil}

\author[0000-0003-4418-4289]{J. C. N. de Araujo}
\affiliation{Divis\~ao de Astrof\'{\i}sica, Instituto Nacional de Pesquisas Espaciais, Avenida dos Astronautas 1758, 12227--010, S\~ao Jos\'e dos Campos, SP, Brazil}

\author[0000-0002-7470-5703]{S. O. Kepler}
\affiliation{Instituto de F\'{\i}sica, Universidade Federal do Rio Grande do Sul, 91501-970 Porto Alegre, RS, Brazil}

\author[0000-0003-4904-0014]{J. A. Rueda}
\affiliation{ICRANet, Piazza della Repubblica 10, I-65122 Pescara, Italy}
\affiliation{ICRANet-Ferrara, Dip. di Fisica e Scienze della Terra, Universit\`a degli Studi di Ferrara, Via Saragat 1, I--44122 Ferrara, Italy}
\affiliation{Dip. di Fisica e Scienze della Terra, Universit\`a degli Studi di Ferrara, Via Saragat 1, I--44122 Ferrara, Italy}
\affiliation{INAF, Istituto di Astrofisica e Planetologia Spaziali, Via Fosso del Cavaliere 100, 00133 Rome, Italy}

\begin{abstract}
Double white dwarf (DWD) mergers are possibly the leading formation channel of massive, rapidly rotating, high-field magnetic white dwarfs (HFMWDs). However, the direct link connecting a DWD merger to any observed HFMWD is still missing. We here show that the HFMWDs SDSS~J221141.80+113604.4 (hereafter J2211+1136) and ZTF~J190132.9+145808.7 (hereafter J1901+1458), might be DWD merger products. J2211+1136 is a $1.27\, M_\odot$ WD with a rotation period of $70.32$~s and a surface magnetic field of $15$~MG. J1901+1458 is a $1.327$--$1.365\, M_\odot$ WD with a rotation period of $416.20$~s, and a surface magnetic field in the range $600$--$900$~MG. With the assumption of single-star evolution, the currently measured WD masses and surface temperatures, the cooling ages of J2211+1136 and J1901+1458 are, respectively, $2.61$--$2.85$ Gyr and $10$--$100$~Myr. We hypothesize that these WDs are DWD merger products and compute the evolution of the post-merged configuration formed by a central WD surrounded by a disk. We show that the post-merger system evolves through three phases depending on whether accretion, mass ejection (propeller), or magnetic braking dominates the torque onto the central WD. We calculate the time the WD spends in each of these phases and obtain the accretion rate and disk mass for which the WD rotational age, i.e., the total time elapsed since the merger to the instant where the WD central remnant reaches the current measured rotation period, agrees with the estimated WD cooling age. We infer the mass values of the primary and secondary WD components of the DWD merger that lead to a post-merger evolution consistent with the observations.

\end{abstract}



\section{Introduction} \label{sec:1}

It has been proposed for a long time that double white dwarf (DWD) mergers can produce high-field magnetic white dwarfs (HFMWDs; see, e.g., \citealp{2000PASP..112..873W}). There are increasing observational results pointing to this scenario (see below), but still, there is a missing direct link connecting a DWD merger to any observed HFMWD. Here, we aim to provide such a link.

In general, the central remnant of a DWD merger can be ($i$) a stable newborn WD, ($ii$) a type Ia supernova (SN), or ($iii$) a newborn neutron star (NS). Sub-Chandrasekhar remnants might end either as ($i$) or ($ii$) \citep{1990ApJ...348..647B, 2012ApJ...746...62R, 2013ApJ...767..164Z, 2018ApJ...857..134B, 2019MNRAS.487..812B, 2021ApJ...906...53S}, super-Chandrasekhar remnants as ($ii$) or ($iii$) \citep{1985A&A...150L..21S, 2016MNRAS.463.3461S, 2018ApJ...857..134B, 2019MNRAS.487..812B, 2021ApJ...906...53S}. We are here interested in those DWD mergers whose remnant is a stable, ultramassive ($\gtrsim 1\,M_\odot$ but sub-Chandrasekhar) HFMWD. Those central remnants might avoid unstable burning leading to an SN~Ia if their central density remains under some critical value of a few $10^9$~$\mathrm{g\,cm^{-3}}$ \citep[see][for further details]{2018ApJ...857..134B}. 
Numerical simulations show that HFMWDs might indeed originate in DWD mergers (see, e.g., \citealp{2012ApJ...749...25G}, and Sec. \ref{sec:2} for further details). The general merged configuration is a central remnant that contains the mass of the (undisrupted) primary, surrounded by a hot corona with about half of the (disrupted) secondary mass, and a rapidly rotating Keplerian disk with roughly the other half of the secondary mass. Little mass ($\sim 10^{-3}\, M_\odot$) is ejected from the system (see Sec.~\ref{sec:4} for further details). The hot and convective corona works as an efficient $\alpha\omega$-dynamo that might lead to magnetic fields of $\lesssim 10^{10}$ G \citep{2012ApJ...749...25G}. For a recent discussion of the emergence of high magnetic fields in observed WDs, we refer the reader to \citet{2022arXiv220802655B}.

Therefore, theory tells us that stable WD remnants of DWD mergers can exist \citep[see, e.g.,][]{2021ApJ...906...53S}. It remains to have observational support. Observations confirm the existence of HFMWDs with magnetic field strengths in the range $10^6$--$10^9$ G \citep{2009A&A...506.1341K, 2015SSRv..191..111F, 2016MNRAS.455.3413K}, and that most of them are massive \citep[see, e.g.,][]{2016MNRAS.455.3413K}. The latest measurements of the transverse velocities of massive WDs also suggest that a fraction of them are DWD merger products \citep[see][and references therein]{2020ApJ...891..160C}. There is an additional observational argument supporting this conclusion. The rate of DWD mergers is expected to be $(5$--$9)\times 10^{-13}$~yr$^{-1}$~$M_\odot^{-1}$ \citep{2017MNRAS.467.1414M, 2018MNRAS.476.2584M}. Using a Milky Way-like stellar mass $6.4\times 10^{10}~M_\odot$ and the extrapolating factor of Milky Way equivalent galaxies, $0.016$~Mpc$^{-3}$ \citep{2001ApJ...556..340K}, the above rate translates into a local cosmic merger rate of $(3.7$--$6.7)\times 10^5$~Gpc$^{-3}$~yr$^{-1}$. This merger rate is $5$--$8$ times larger than the population of SN~Ia \citep[see, e.g.,][]{2009ApJ...699.2026R}. Therefore, even if we were to explain the entire SN Ia population with the double-degenerate channel, i.e., with DWD mergers, we can safely conclude that many DWD mergers do not produce SNe~Ia \citep[see, also,][]{ 2020ApJ...891..160C}.

All the above leads to a rather obvious question (but with no obvious answer), where and which are the WDs produced by (some of) those mergers? To answer this question, we here reconstruct the DWD progenitor of two recently detected HFMWDs, SDSS~J221141.80+113604.4 (hereafter J2211+1136; \citealp{2021ApJ...923L...6K}) and ZTF~J190132.9+145808.7 (hereafter J1901+1458; \citealp{2021Natur.595...39C}). J2211+1136 has a mass of $1.27\,M_\odot$, rotation period of $70.32$~s, surface magnetic field strength of $15$~MG, effective temperature $T_{\rm eff} \approx 9020$~K, and cooling age $2.61$--$2.85$ Gyr \citep{2021ApJ...923L...6K, 2021MNRAS.503.5397K}. The corresponding parameters of  J1901+1458 are a mass in the range of $1.327$--$1.365\, M_\odot$, rotation period of $416.20$~s, surface magnetic field in the range $600$--$900$~MG, effective temperature $T_{\rm eff} = 46000^{+19000}_{-8000}$~K, and cooling age $10$--$100$~Myr \citep{2021Natur.595...39C}. The cooling age is estimated assuming single-star evolution and the currently measured WD masses and effective temperatures.

Following numerical simulations of DWD mergers, we model the post-merger configuration as a central HFMWD remnant surrounded by a Keplerian disk (see Sec.~\ref{sec:2} for details). We compute the post-merger rotational evolution of the system and infer the model parameters for which the WD rotational age, i.e., the time at which it reaches the current value of the rotation period, agrees with the estimated cooling age. We show that the post-merger configuration evolves through three phases dominated by accretion, mass ejection (propeller), and magnetic braking (see Sec.~\ref{sec:3} for details). The latter phase dominates the duration of the rotational age. We derive the accretion rate, the disk mass, and the mass of the pre-merger DWD primary and secondary binary components for which the post-merger system agrees with observations.

We organize the article as follows. In Sec.~\ref{sec:2}, we summarize the properties of the merged configuration derived from numerical simulations that serve as the starting point of our calculations. Section~\ref{sec:3} describes the theoretical treatment to compute the rotational evolution of the post-merger configuration. Section~\ref{sec:4} explains the different types of rotational evolution. We apply in Sec.~\ref{sec:5} the theoretical model to the HFMWDs J2211+1136 and J1901+1458. In Sec. \ref{sec:6}, we infer the parameters of the pre-merger DWD progenitors for the two sources. Finally, Sec.~\ref{sec:7} outlines the conclusions of this article.

\section{Merger and Post-Merger Properties} \label{sec:2}

According to numerical simulations of DWD mergers \citep[see, e.g.,][]{1990ApJ...348..647B, 2004A&A...413..257G, 2009A&A...500.1193L, 2012A&A...542A.117L, 2012ApJ...746...62R,
2013ApJ...767..164Z, 2014MNRAS.438...14D, 2018ApJ...857..134B}, the merged configuration is the central, rigidly rotating, isothermal WD core of mass $M_{\rm core}$, surrounded by a hot envelope of mass $M_{\rm env}$ with differential rotation and a rapidly rotating Keplerian disk of mass $M_d$. The mass of the disrupted secondary star distributes between the envelope and the disk. Some material of mass $M_{\rm fb}$ falls back onto the WD remnant, and only a tiny amount of mass $M_{\rm ej}$ escapes from the system. \citet{2014MNRAS.438...14D} obtained the following fitting polynomials of the properties of the merged configurations from numerical simulations of DWD mergers for a variety of initial conditions
\begin{subequations}
\begin{align}
M_{\rm core} &= M_{\rm tot} (0.7786 - 0.5114\,q), \label{eq:mcore}\\
M_{\rm env} &= M_{\rm tot} (0.2779 - 0.464\,q + 0.7161\,q^2), \label{eq:menv}\\
M_d &= M_{\rm tot} (-0.1185 + 0.9763\,q - 0.6559\,q^2), \label{eq:md}\\
M_{\rm fb} &= M_{\rm tot} (0.07064 - 0.0648\,q), \label{eq:mfb}\\
M_{\rm ej} &= \frac{0.0001807 M_{\rm tot}}{-0.01672 + 0.2463\,q - 0.6982\,q^2 + q^3},\label{eq:mej}
\end{align}
\end{subequations}
where $M_{\rm tot} = M_1 + M_2$ is the total binary mass, with $M_1$ and $M_2$ the masses of the primary and secondary, and $q \equiv M_2/M_1 \leq 1$ is the binary mass ratio. The goodness of the polynomial fitting was reported in \citet{2014MNRAS.438...14D} with $R^2$ statistic values of $0.97$, $0.88$, $0.78$ and $0.8$, respectively for the first four fitting functions (\ref{eq:mcore})--(\ref{eq:mfb}), which means they fit the $97\%$, $88\%$, $78\%$, and $80\%$ of the corresponding variance.

We model the post-merger evolution after the short-lived phase in which the WD core incorporates the envelope. The post-merger system is thus composed of the WD remnant of mass $M$, radius $R$, surrounded by the accretion disk of mass $M_d$. As for the magnetic field configuration, we adopt a dipole+quadrupole model with a dipole strength $B$ and a quadrupole strength $B_{\rm quad}$. Hence, the magnetic dipole moment is $\mu = B R^3$.

For fixed $M_{\rm tot}$, Eq. (\ref{eq:mej}) shows that the unbound mass decreases for increasing $q$, so the lowest value is obtained for the largest possible binary mass-ratio ($q=1$), i.e., $M_{\rm ej} \approx 3.4\times 10^{-4} M_{\rm tot}$. Given the approximate mass conservation, we estimate the mass of the final WD by

\begin{equation}\label{eq:M1pieces}
    M = M_{\rm core} + M_{\rm env} + M_{\rm fb} + M_{\rm acc},
\end{equation}
where $M_{\rm acc} \leq M_d$ is the accreted mass. As we shall show, $M < M_{\rm tot}$ because some mass is ejected from the system during the propeller phase (see more details in Sec.~\ref{sec:4}). With the aid of Eqs. (\ref{eq:mcore})--(\ref{eq:mej}), Eq. (\ref{eq:M1pieces}), and the estimate of the accreted mass, in Sec. \ref{sec:6} we estimate the parameters of the merging binary.  

In the following calculations, we assume a constant value of $M$ given by the current value of the mass of the WD, i.e., the final value given by Eq.~(\ref{eq:M1pieces}), neglecting the effect of the increase in mass, $M_{\rm acc}$. We also assume a constant accretion rate onto the central WD, $\dot{M}$. These assumptions have a negligible effect on the results because mass accretion and ejection consume the disk mass in a timescale much shorter than the WD lifetime, and so magnetic braking dominates the rotational evolution.

\section{Post-Merger Rotational Evolution} \label{sec:3}

\subsection{Accretion and propeller torque}

When the magnetosphere radius, $R_m$, is smaller than the WD radius, $R$, the disk extends up to the WD surface, i.e., $r_i=R$. When $R_m > R$, the disk extends up to $r_i =R_m$. Therefore, we have
\begin{equation}\label{eq:ri}
    r_i = {\rm max}(R,R_m),
\end{equation}
where $R_m$ is the Alfvén radius\footnote{This magnetosphere radius assumes spherical accretion and a pure dipole magnetic field configuration.} \citep[see, e.g.,][]{1972A&A....21....1P}
\begin{equation}\label{eq:Rm}
    R_m = \left( \frac{\mu^2}{\dot{M} \sqrt{2 G M}}  \right)^{2/7}.
\end{equation}

From Eq.~(\ref{eq:Rm}), we obtain that the condition $R_m > R$ is satisfied for accretion rates
\begin{equation}\label{eq:cond1}
    \frac{\dot{M}}{M_\odot\,{\rm yr}^{-1}} < 9.74 \times 10^{-4} \frac{B_{8}^2\,R_8^{5/2}}{\left( M/M_\odot\right)^{1/2}},
\end{equation}
where $B_{8} = B/(10^8\,\rm G)$ and $R_8 = R/(10^8\,\rm cm)$.

As we shall see, the accretion rate onto the remnant WD is much lower than the above value, so the WD evolves always in the regime $R_m>R$. In this case, the magnetic field lines thread the disk at $r_i = R_m$ and the matter flow from the disk to the WD through the magnetic field lines. Whether this mass flow spins up or down the central WD depends on the value of the so-called \textit{fastness} parameter
\begin{equation}\label{eq:fastness}
    \omega \equiv \frac{\Omega}{\Omega_K},
\end{equation}
where $\Omega_K$ is the Keplerian angular velocity at $r_i = R_m$
\begin{equation}\label{eq:OmK}
    \Omega_K = \sqrt{\frac{G M}{R_m^3}}.
\end{equation}

The specific (i.e., per unit mass) angular momentum of the matter leaving the disk is $l_i = \Omega_K r_i^2 = \Omega_K R_m^2 = \sqrt{G M R_m}$, while the specific angular momentum of the co-rotating magnetosphere at $r=R_m$ is $l_m = \Omega R_m^2$. Therefore, the WD will change its angular momentum at a rate given by \citep[][]{1999ApJ...520..276M}
\begin{equation}\label{eq:Tacc}
    \dot{J}_{\rm acc} = T_{\rm acc} = (l_i - l_m)\dot{M} = \delta \left(1-\omega \right),
\end{equation}
where
\begin{equation}
    \delta \equiv \dot{M} R_m^2\,\Omega_{K} = \dot{M} \sqrt{G M R_m},
\end{equation}
being $\dot{M}$ the accretion rate, i.e., the rate at which mass flows from the disk to the WD at the inner disk radius, $r=r_i=R_m$. When $\omega < 1$, the inflowing material accretes onto the WD and transfers angular momentum to it (exerts a positive torque). When $\omega >1$, the system enters into the so-called \textit{propeller} regime in which the WD centrifugal barrier expels the inflowing mass from the disk. Such mass leaves the system removing angular momentum from the central WD, i.e., it exerts a negative torque onto it. For additional discussions about the propeller mechanism, we refer the reader, e.g., to \citet{1975A&A....39..185I} and \citet{1995ApJ...449L.153W}.

\subsection{Magnetic dipole torque}

The central remnant is also subjected to the torque by the magnetic field. Since the ratio between the stellar radius, $R$, and the light-cylinder radius, $R_{\rm lc} = c/\Omega$, is small, i.e., $R/R_{\rm lc} = \Omega R/c \lesssim 10^{-3}$, finite-size effects in the determination of the radiation field can be safely neglected. Therefore, we use the torque exerted by a point-like dipole+quadrupole magnetic field configuration \citep{2015MNRAS.450..714P}
\begin{align}
    T_{\rm mag} &= T_{\rm dip} + T_{\rm quad},\label{eq:Tmag}\\
    T_{\rm dip} &= -k_{\rm dip}\,\Omega^3,\label{eq:Tdip}\\
    T_{\rm quad} &= -k_{\rm quad}\,\Omega^5,\label{eq:Tquad}
\end{align}
where
\begin{align}
     k_{\rm dip} &=\frac{2}{3} \frac{B^2 R^6}{c^3} \sin^2\theta,\label{eq:kdip}\\
     k_{\rm quad} &= \frac{32}{135} \frac{B_{\rm quad}^2 R^8 }{c^5} \sin^2\theta_1 (\cos^2\theta_2+10\sin^2\theta_2),\label{eq:kquad}
\end{align}
being $\theta$ the inclination angle of the magnetic dipole moment with respect to the WD rotation axis, and the angles $\theta_1$ and $\theta_2$ specify the geometry of the quadrupole field. We can write the total magnetic torque as
\begin{equation}\label{eq:Tmag01}
    T_{\rm mag} = -\frac{2}{3} \frac{B^{2} R^{6} \Omega^{3}}{c^{3}} \left ( \sin^{2}\theta + \eta^{2}\frac{16}{45} \frac{R^{2} \Omega^{2}}{c^{2}} \right ),
\end{equation}
where $\eta$ is a parameter that measures the quadrupole to dipole magnetic field strength ratio as
\begin{equation}\label{eq:eta}
    \eta = \frac{B_{\rm quad}}{B} \sqrt{\sin^{2}\theta_{1}(\cos^{2}\theta_{2} + 10\sin^{2}\theta_{2})} .
\end{equation}

Having defined the torques acting onto the central WD, we can write the equation of angular momentum conservation as
\begin{equation}\label{Eq:Ttot}
    T_{\rm tot} = T_{\rm acc} + T_{\rm mag} = \dot{J} \approx I \dot{\Omega},
\end{equation}
whose integration gives the evolution of the WD rotational properties (e.g., angular momentum and angular velocity). The last equality neglects the change in time of the WD moment of inertia, $I$, as required by self-consistency with the approximation of constant mass. The above differential equation can be integrated given initial condition to the angular velocity, $\Omega_0 = \Omega (t_0=0)$, and setting all the model parameters, i.e., $\{M, R, I, \dot{M}, M_d, B, B_{\rm quad}, \theta, \theta_1, \theta_2,\}$. The qualitative and quantitative features of the result are not sensitive to the initial condition of the angular velocity, $\Omega_0$. We shall explore a variety of initial rotation periods ranging from a few seconds to hundreds of seconds.

The magnetic field and the mass of the WD are set by observations (and so its radius and moment of inertia from its structure properties, e.g., mass-radius relation), so it remains to set $\dot{M}$, $M_d$ and $\theta$. Without loss of generality, we shall assume an orthogonal dipole, $\theta = 90^\circ$, and an $m = 1$ mode for the quadrupole, i.e., $(\theta_1,\theta_2) = (\pi/2,0)$. For the disk mass, we shall set values around $M_d \approx 0.30~M_\odot$ according to numerical simulations \citep[see, e.g.,][and Secs. \ref{sec:5} and \ref{sec:6} for details on the effect of different disk masses]{2014MNRAS.438...14D, 2018ApJ...857..134B}. Therefore, it remains only to set the value of $\dot{M}$. We shall do it by requesting that the value of $\Omega$ equals the current observed value at the evolution time consistent with estimated WD cooling age.

\section{Types of rotational evolution}\label{sec:4}

The WD might follow different types of rotational evolution depending on the model parameters and initial conditions. In the most general case, the system evolves through three stages until it reaches the current rotation period: a \textit{phase I} of accretion (the WD spins up) or ejection of matter (the WD spins down) by propeller, a \textit{phase II} in which accretion and matter ejection episodes balance each other so the WD spin remains at an equilibrium value, and the \textit{phase III} in which the system enters once the disk mass ends, so the WD spins down because of magnetic braking.

Because DWD mergers always form a debris disk around the newborn central remnant \citep[see, e.g.,][]{2009A&A...500.1193L, 2014MNRAS.438...14D}, we rule out an evolution with only phase III, i.e., without either accretion or matter ejection and only evolving due to magnetic braking. Therefore, the post-merger WD necessarily starts its evolution either at the phase I or II. 

\begin{figure}
\includegraphics[width=\hsize,clip]{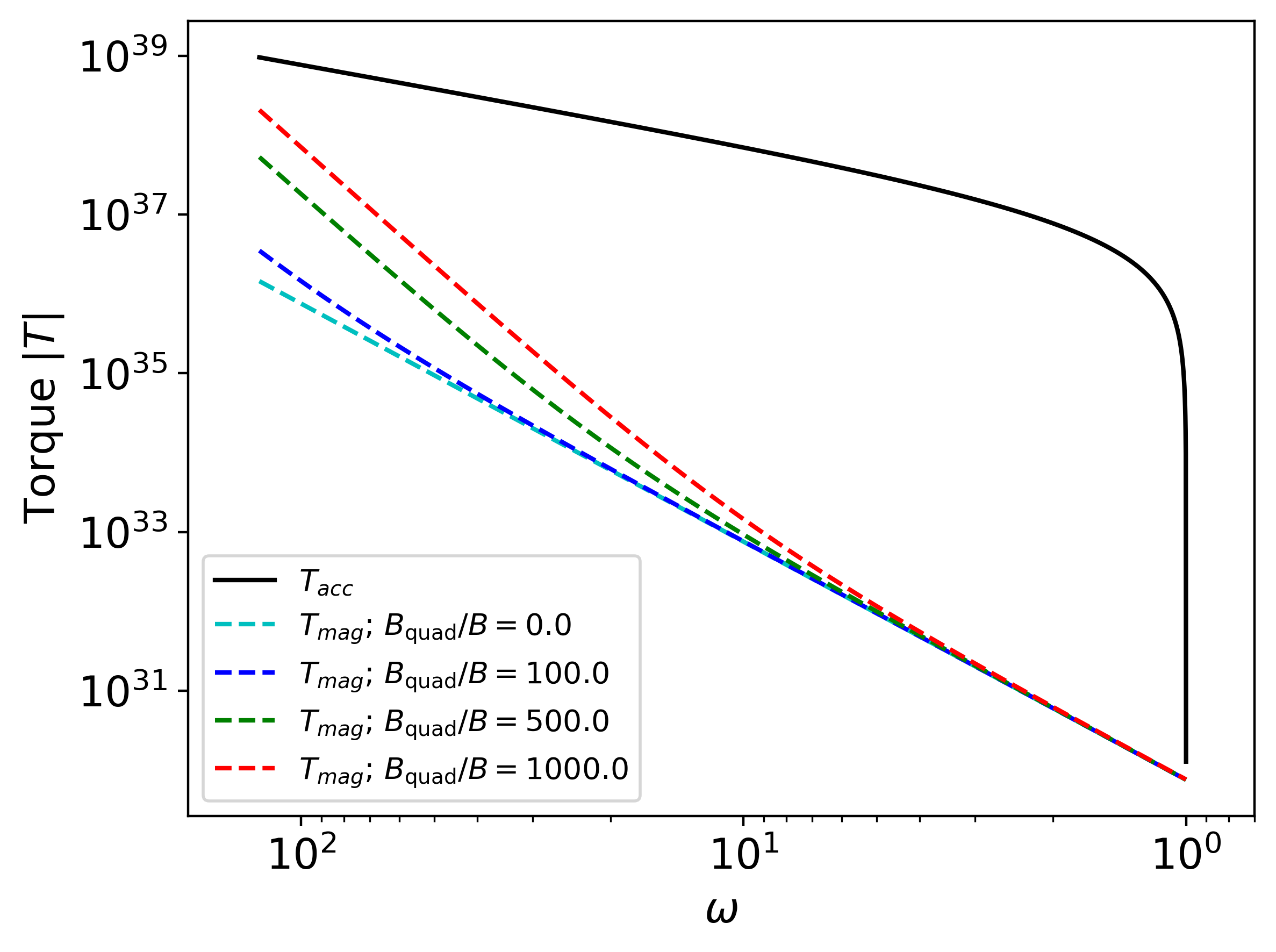}
\caption{$|T_{\rm acc}|$ and $|T_{\rm mag}|$ as a function of $\omega >1$ where we considered a WD with $B = 100$ MG, $\dot{M} = 10^{-7} M_\odot$ yr$^{-1}$ and the quadrupole to dipole magnetic field strength ratio $B_{\rm quad}/B$. The ratio $B_{\rm quad}/B = 0.0$ correspond to the case of a pure dipole magnetic field.}\label{fig:wfXT}
\end{figure}

\begin{figure}
\includegraphics[width=\hsize,clip]{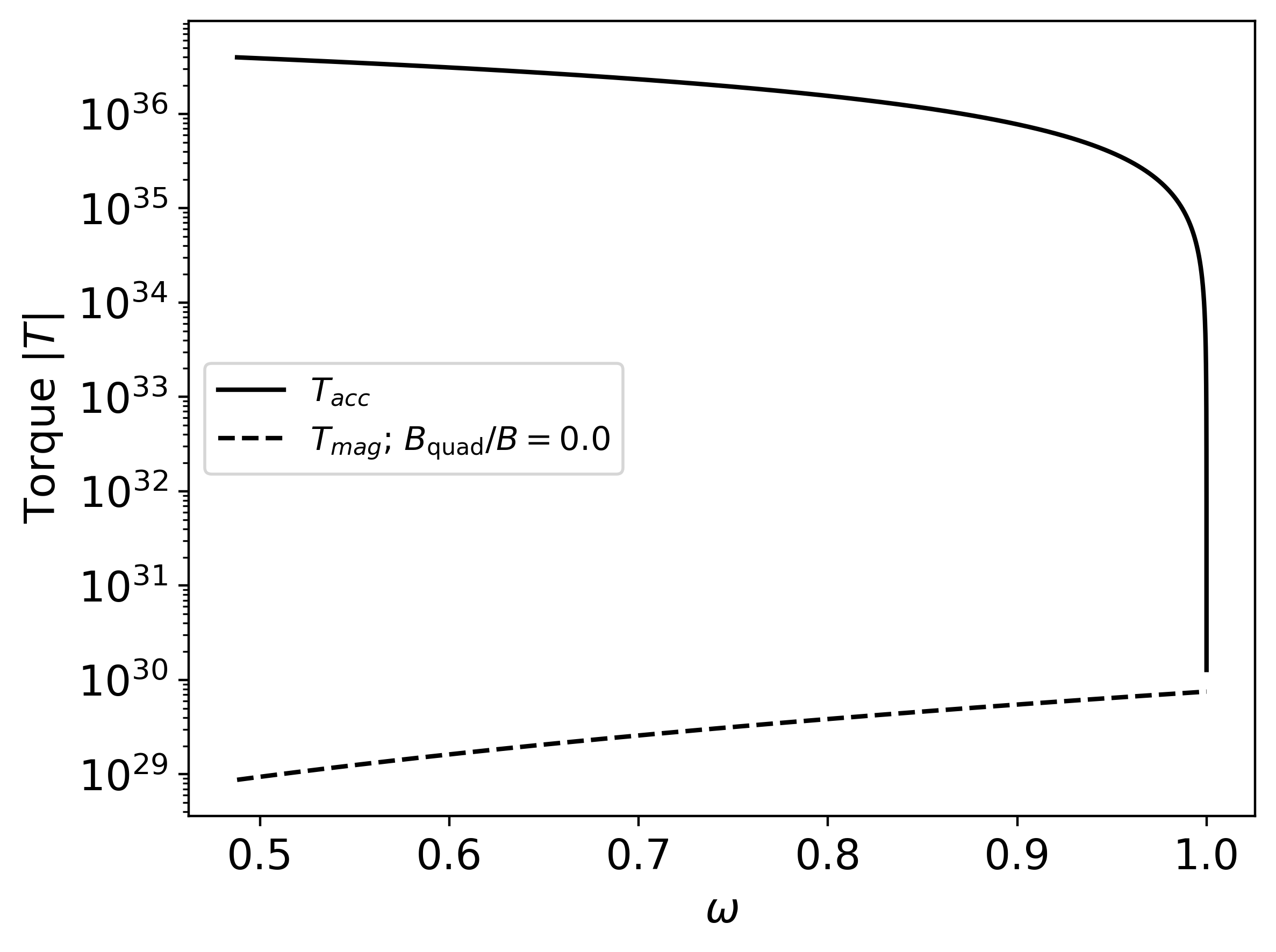}
\caption{$|T_{\rm acc}|$ and $|T_{\rm mag}|$ as a function of $\omega <1$ where we considered a WD with $B = 100$ MG, $\dot{M} = 10^{-7} M_\odot$ yr$^{-1}$ and the quadrupole to dipole magnetic field strength ratio $B_{\rm quad}/B = 0.0$, since the effect of the quadrupole magnetic field is very small when we take into account the values of $\omega <1$ presented here (See Eq.\ref{eq:Tmag01}).}\label{fig:wfXT02}
\end{figure}

We now focus on the most general case, i.e., when the system evolves through the above three stages until it reaches the current rotation period. The division of the evolution into three phases depends on the value of the fastness parameter, that is, $\omega >1$, $\omega \approx 1$ and $\omega <1$. Depending on the initial angular velocity, $\Omega_0$, the initial value of the fastness parameter can be either $\omega_0 >1$ ($\Omega_0 > \Omega_K$) or $\omega_0 <1$ ($\Omega_0 < \Omega_K$). The torque $T_{\rm acc}$ is either negative when the propeller mechanism is active ($\omega >1$) or positive when the accretion process is active ($\omega <1$). On the other hand, the magnetic torque $T_{\rm mag}$ always removes angular momentum. When there is mass flowing from the inner radius of the disk, the propeller or the accretion dominates the changes in the rotational period given that $T_{\rm acc}$ dominates over $T_{\rm mag}$. Figure \ref{fig:wfXT} shows $T_{\rm acc}$ and $T_{\rm mag}$ as a function of $\omega >1$ up to the value of $\omega \approx 1$, for fiducial values of the dipole magnetic field and the accretion rate, respectively, $B = 100$ MG and $\dot{M} = 10^{-7} M_\odot$ yr$^{-1}$. We consider $T_{\rm mag}$ with different values of $B_{\rm quad}/B$ in order to assess the effect of $B_{\rm quad}$ on the spin evolution. Figure \ref{fig:wfXT02} is analogous to Fig. \ref{fig:wfXT}, but for $\omega <1$. Here, we only consider the pure magnetic dipole case, $B_{\rm quad}/B = 0$, because for low values of the angular velocity the torque by the quadrupole magnetic field is very small compared to that produced by the magnetic dipole [see Eq. (\ref{eq:Tmag01})]. These figures shows that $|T_{\rm acc}| \gg |T_{\rm dip}|$ for the entire present phase, with the only exception when $\omega\approx 1$ where they become comparable as $T_{\rm acc}$ drops significantly.

It is worth mentioning that for $B_{\rm quad}/B = 1000$, the intensity of $T_{\rm mag}$ approaches the intensity of $T_{\rm acc}$ for values of the angular velocity around a few seconds. However, when the angular velocity decreases, $T_{\rm mag}$ drops rapidly while $T_{\rm acc}$ remains high for nearly the entire phase of $\omega >1$. Thus, $T_{\rm acc}$ still dominates the torque and $T_{\rm mag}$ (even with $B_{\rm quad}/B = 1000$) contributes very little to this first evolution stage.

Taking the above into account in this first regime, we can approximate with sufficient accuracy the total torque by $T_{\rm tot} \approx T_{\rm acc}$. Within this approximation, Eq. (\ref{Eq:Ttot}) has the analytic solution
\begin{equation}\label{eq:omega1}
    \omega =  1 + (\omega_{0} - 1)\,e^{-t/\tau_{\rm acc}},
\end{equation}
where $\omega_0 = \Omega_0/\Omega_K$ is initial fastness parameter and $\tau_{\rm acc}$ is the timescale (e-folding time) of the propeller/accretion phase
\begin{equation}
    \tau_{\rm acc} = \frac{I\,\Omega_K}{\delta} = \frac{I}{\dot{M} R_m^2} = 50.27 \frac{I_{49}}{\dot{M}_{-7} R_{m,9}^2}\,\,{\rm kyr},
\end{equation}
where $I_{49}$ is the moment of inertia in units of $10^{49}$ g cm$^2$, $\dot{M}_{-7}$ is the accretion rate in units of $10^{-7} M_\odot$ yr$^{-1}$ and $R_{m,9}$ is the Alfvén radius in units of $10^9$ cm.

When $\omega \approx 1$, the WD enters the phase II of evolution, characterized by $T_{\rm acc} \approx T_{\rm mag}$. When this occurs, $T_{\rm mag}$ decelerates the star to an angular velocity slightly smaller than $\Omega_K$, so to a fastness parameter $\omega\lesssim 1$. Then, $T_{\rm acc}$ turns positive, the WD accretes matter, spinning it up. This happens until the WD returns once more to the regime of $\omega \gtrsim 1$. Upon reaching this regime, due to the torques $T_{\rm acc}$ and $T_{\rm mag}$, the WD rotation decelerates once more until $\omega \lesssim 1$. The accretion acts, and the propeller process ceases again. Summarizing, in this stage the WD goes through a continuous spin-up and spin-down stages in which the fastness parameter oscillates around unity, so the angular velocity oscillates around an \textit{equilibrium} value, $\Omega \approx \Omega_{\rm eq} = \Omega_K$. Therefore, we can assume that in this phase $\Omega$ remains constant at
\begin{align}\label{eq:Omeeq}
    \Omega_{\rm eq} &= \Omega_K = \left [ \frac{\sqrt{2} \, (GM)^{5/3} \, \dot{M}}{B^{2} \, R^{6}} \right ]^{3/7}\nonumber \\
    &= 0.225 \left [ \frac{(M/M_{\odot })^{5/3} \, \dot{M}_{-7}}{B_8^2 \, R_8^6} \right ]^{3/7}\,\, {\rm rad \, s^{-1}}.
\end{align}

This phase lasts until the disk can feed the accretion and propeller. Thus, the duration timescale of this phase is of the order of $\tau_{\rm disk} \approx M_d/\dot{M} \sim 10^6$ yr, that is the time required to consume the disk mass. 

After the disk is exhausted, the system evolves to the regime $\omega < 1$. Without mass flowing from the disk, only the magnetic dipole exerts torque. Equation (\ref{eq:Tmag01}) and Fig. \ref{fig:wfXT} show that the effect of the quadrupole magnetic field on the magnetic torque is negligible in the range of angular velocities of the regime $\omega \lesssim 1$. In this case, the torque due to dipole radiation dominates and we can accurately approximate the total torque by $T_{\rm tot} = T_{\rm mag} \approx T_{\rm dip}$. Thus, we can solve Eq. (\ref{Eq:Ttot}) analytically
\begin{equation}
    \omega = \left ( 1 + \frac{\Delta t}{\tau_{\rm dip}} \right )^{-1/2},
     \label{Eq:Omedip}
\end{equation}
where $\Delta t = t-t_i$, being $t_i$ the initial time of the pure magnetic dipole torque phase, and we have used that this phase starts with an initial value of the fastness parameter equal to unity, i.e., $\omega(t=t_i)=1$. The characteristic spindown timescale $\tau_{\rm dip}$ is given by
\begin{align}\label{eq:taudip}
    \tau_{\rm dip} &= \frac{I}{2\,k_{\rm dip}\,\Omega_K^2} = \frac{c^{3} I R_m^3}{2 G M B^2 R^3 (1+\sin^2\theta)}\nonumber \\
    &= 3.22 \frac{I_{49} R_{m,9}^3}{(M/M_\odot) B_8^2 R_8^3 (1+\sin^2\theta)}\,\, {\rm Gyr},
\end{align}
which is much longer than the timescale of the previous phases. The above implies that we can approximate the total post-merger age of the WD to the time it spends in this final phase.

We can invert Eq.~(\ref{Eq:Omedip}) and find the elapsed time $\Delta t_{\rm obs}$ for the WD to reach an observed angular velocity, $\Omega_{\rm obs}$, i.e.,
\begin{equation}\label{eq:tobs}
    \Delta t_{\rm obs} = \tau_{\rm dip} \left [\left ( \frac{\Omega_K}{\Omega_{\rm obs}}  \right )^{2} -1 \right ].
\end{equation}

Since $\tau_{\rm dip}$ depends on $\Omega_K$, and the latter depends on $R_m$, and so on $\dot{M}$, we can use Eq. (\ref{eq:tobs}) to express $\dot{M}$ in terms of $\Omega_{\rm obs}$ and $\Delta t_{\rm obs}$
\begin{equation}\label{eq:Mdot}
    \dot{M} = \frac{B^2 R^6 \, \Omega_{\rm obs}^{7/3}}{\sqrt{2} \, (GM)^{5/3} } \left( 1 - \frac{2\,k_{\rm dip}\,\Omega_{\rm obs}^2}{I} \Delta t_{\rm obs} \right )^{-7/6}.
\end{equation}

Therefore, Eq. (\ref{eq:Mdot}) allows to obtain, from observational parameters such as mass, angular velocity, magnetic field and the estimated age of the WD (e.g., the cooling age), the accretion rate for which the rotational evolution agrees with observations.

\section{Analysis of specific sources} \label{sec:5}

Having described all the generalities of the post-merger evolution, we turn to describe the rotational evolution of two observed HFMWDs, namely, J2211+1136 and J1901+1458. We infer the time the WD spends in each phase, and calculate the accretion rate that leads to the rotational evolution to agree with the estimated WD cooling age. This assumption agrees with the fact that the cooling tracks of these sources are estimated considering the current mass of the WD, so the cooling age is the evolution time of the post-merger central remnant after the envelope has been fully incorporated into the isothermal core and the WD composition has settled down \citep[see e.g.][for more details]{2021ApJ...916..119S}. The above is also supported by the fact that the initial phase of fusion containing the envelope is short-lived ($\sim 10^4$--$10^5$ yr; see, e.g., \citealp{2012ApJ...749...25G, 2021ApJ...906...53S}) compared to the estimated age of the WD and, as we show in this article, also the duration of the accreting phase is negligible. 

We do not take into account any delay due to crystallization, phase separation due to sinking or dilution of heavier elements, or nuclear energy that can possibly occur as proposed e.g. by \citet{Cheng20,Blouin21}.

\subsection{Rotational evolution of J2211+1136} \label{subsec:SDSS}

J2211+1136 is a recently observed isolated WD with a rotation period $P_{\rm obs} = 70.32$ s \citep{2021ApJ...923L...6K}. It has a mass $M=1.27 M_{\odot}$, a stellar radius $R=3210$ km,\footnote{We estimate the radius from the measured mass and surface gravity, $\log(g) = 9.214$ \citep{2021ApJ...923L...6K} i.e., $R = \sqrt{G M/g}$.} and a surface (dipole) magnetic field $B=15$ MG. The cooling age is in the range $t_{\rm cool} = 2.61$--$2.85$ Gyr depending on the WD interior composition \citep{2021ApJ...923L...6K, 2021MNRAS.503.5397K}.

\begin{figure}
\includegraphics[width=\hsize,clip]{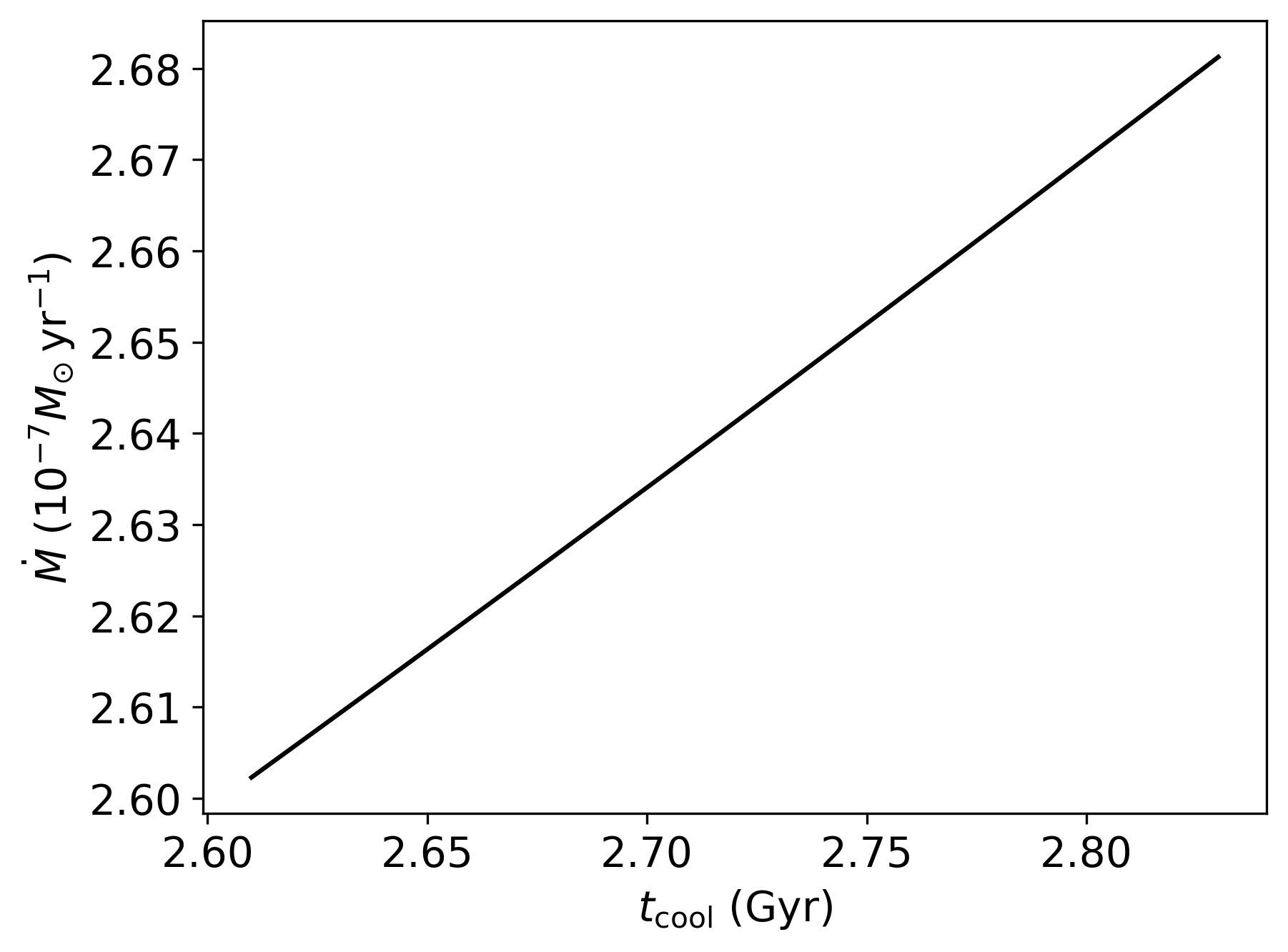}
\caption{Accretion rate as a function of cooling age, calculated from Eq.~(\ref{eq:Mdot}) for J2211+1136. The WD parameters are $M = 1.27 M_{\odot}$, $R = 3.21 \times 10^{8}$ cm, $P_{\rm obs} = 70.3$ s. The dipole magnetic field is $B = 15$ MG and $\theta = 90^{\circ}$.} \label{fig:SDSS_ACC_time}
\end{figure}

First, to explore the evolutionary path of the WD rotation, we need to know the accretion rate values for which the rotational evolution time agrees with the cooling age. For this task, we use Eq.~(\ref{eq:Mdot}), assuming $\Delta t_{\rm obs} = t_{\rm cool}$. Figure \ref{fig:SDSS_ACC_time} shows that this condition implies that $\dot{M}$ must be in the range $\approx (2.60$--$2.68) \times 10^{-7} M_{\odot}$ yr$^{-1}$. For an aligned rotator ($\theta = 0$), the corresponding accretion rate range is $(2.21$-- $2.24)\times 10^{-7} M_{\odot}$ yr$^{-1}$.

\begin{figure*}
\includegraphics[width=\hsize,clip]{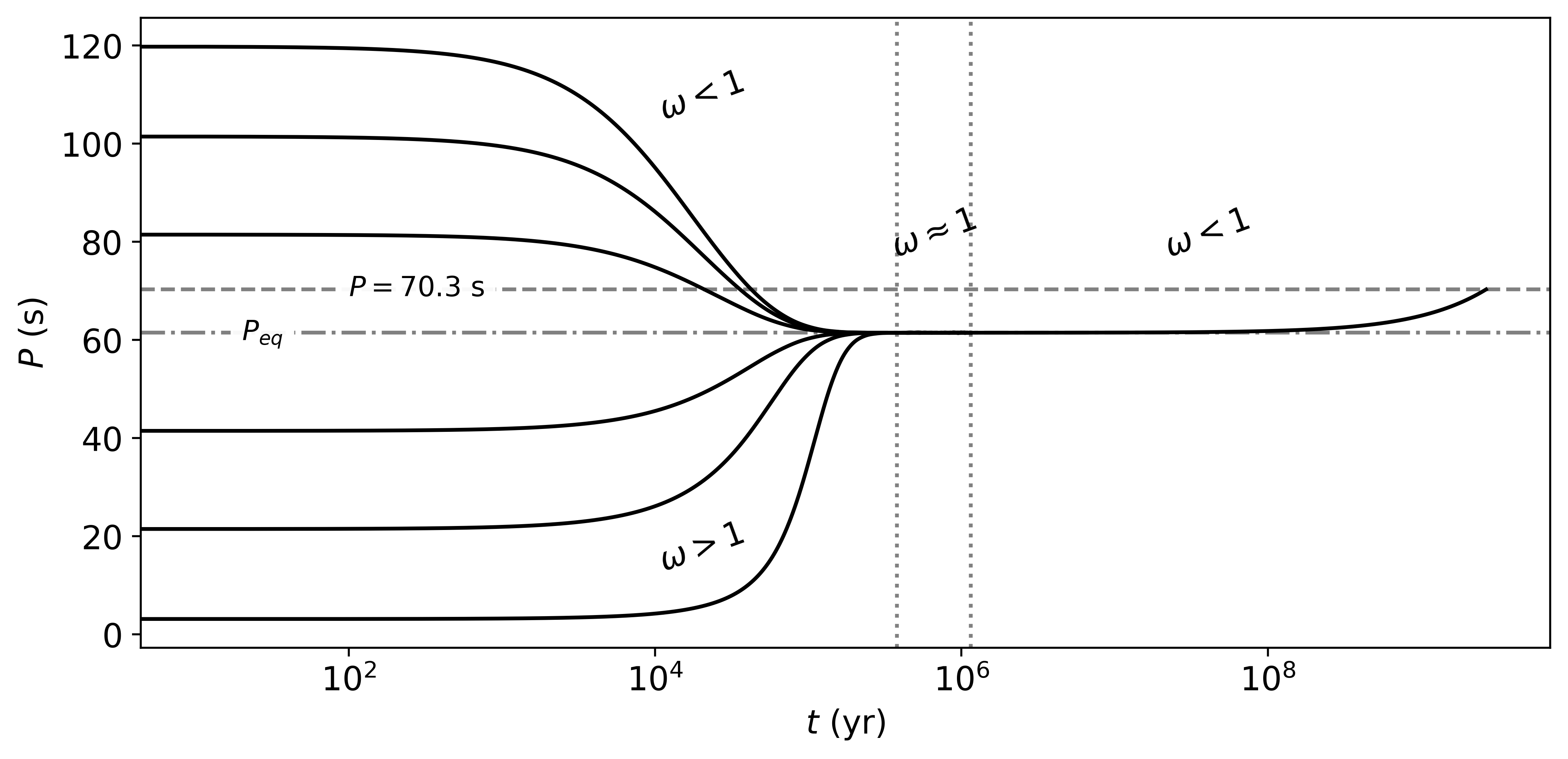}
\caption{Evolution of the rotation period of the J2211+1136 for an accretion rate of $\dot{M} = 2.62 \times 10^{-7} M_{\odot}$ yr$^{-1}$ and for different values of initial rotational periods, $P_{0} = (3.14, \, 21.5, \, 41.5, \, 81.5, \, 101.5, \, 119.8)$ s. The dotted lines divide the evolution into three stages according to the value of $\omega$. In the first stage, WD can start with $\omega > 1$ or $\omega < 1$ depending whether the initial period is below or above the equilibrium period, $P_{\rm eq} = 61.15$ s (Eq. \ref{eq:Omeeq}), respectively. For both values, the involved torques are magnetic torque, $T_{\rm mag}$, and accretion torque, $T_{\rm acc}$. However, $T_{\rm acc}$ is the dominant torque in this phase. For $\omega \approx 1$, $T_{\rm mag} \approx T_{\rm acc}$ and the system goes through the $P_{\rm eq}$. For $\omega < 1$, the involved torque is $T_{\rm mag}$. $T_{\rm acc}$ no longer acts on the star, as in the third stage, the disk has already exhausted. The upper dashed line indicates the current rotation period of the WD.} \label{fig:SDSS_26e8}
\end{figure*}

To analyze in detail the phases of spin evolution, we consider a value of $\dot{M}$ within the above range, e.g., $\dot{M} = 2.62 \times 10^{-7} M_{\odot}$ yr$^{-1}$. Furthermore, we consider six values for the initial rotation period to verify that the solution is not sensitive to this initial condition. For this task, we choose three values below and three values above the equilibrium period, $P_{\rm eq} = 2\pi/\Omega_{\rm eq} \approx 61.5$ s, i.e., $P_{0} = (3.14, \, 21.5, \, 41.5, \, 81.5, \, 101.5, \, 119.8)$ s. Figure \ref{fig:SDSS_26e8} shows the evolution of the WD rotation period for these initial conditions. We observe that irrespective of $P_0$, the WD accelerates or decelerates towards $P_{\rm eq}$ on a comparable timescale. Therefore, the duration of the rotational evolution in the phase I is not sensitive to the specific value of $P_{0}$.

Because of the above result, we examine in detail the evolution curve for a single case, e.g., $P_{0} = 3.14$ s. For the parameters of this WD, and a $M_{d} = 0.30 M_{\odot}$ which lies within the range of consistent values obtained for the disk mass and numerical simulations (see Sec. \ref{sec:6} for more details), the evolution of the rotation period of J2211+1136 crosses the three stages until it reaches the current value of the rotation period. First, it passes through the regime of $\omega > 1$, such that the torques $T_{\rm acc}$ and $T_{\rm dip}$ spindown the WD to a period of $61.5$ s in $\Delta t_1\approx 0.37$ Myr. This time is marked by the first dotted line in Fig. \ref{fig:SDSS_26e8}. The amount of disk mass ejected by the propeller effect during this time is  $M_{\rm loss,1} = \dot{M} \Delta t_1 \approx 0.096 M_{\odot}$.

From this period value, the parameter $\omega \approx 1$ and the system enters the equilibrium period regime, where the WD rotation period is oscillating around $P_{\rm eq}$ (see Eq. \ref{eq:Omeeq}). The system remains at this stage until the disk mass ends. The phase lasts $\Delta t_2\approx 0.78$ Myr. Therewith, adding the duration of the first and second stages, we have up to this point an evolution time of $\Delta t_1+\Delta t_2\approx 1.15$ Myr. This time is marked by the second dotted line in Fig. \ref{fig:SDSS_26e8}. In this phase II, the disk mass loss is divided in equal parts in accretion and ejection, so $M_{\rm acc,2} = M_{\rm loss,2} = (1/2)\dot{M}\Delta t_2 \approx 0.102 M_\odot$. Therefore, by the end of phase II, the total disk mass has been indeed consumed, i.e., $0.096M_\odot+0.204 M_\odot = M_d$. The disk mass has been distributed in a total ejected mass $M_{\rm loss} = M_{\rm loss,1}+M_{\rm loss,2} \approx 0.2 M_\odot$ and a total accreted mass $M_{\rm acc} = M_{\rm acc,2} \approx 0.1 M_\odot$.

After this point, the system enters the regime of $\omega < 1$, where the only active torque is $T_{\rm mag}$. Thus, $T_{\rm mag}$ spins the WD down from a period of $61.5$ s to the observed period $P_{\rm obs} = 70.3$ s, reached at a time of $2.66$ Gyr (see Fig. \ref{fig:SDSS_26e8}). The phase III is by far the longest, so the WD spends most of its evolution in this regime.

In addition to the agreement of the rotational and cooling ages, the full proof of the present scenario would arise from the further agreement of the model spindown rate with the corresponding observational measurement \citep[see, e.g., the case of 4U 0142+61 in][]{2013ApJ...772L..24R}. For the magnetic dipole braking mechanism (see Eq. \ref{eq:Tmag01}), the spindown rate is given by
\begin{equation}
    \dot{P} = \frac{8 \pi^{2}}{3c^{3}} \frac{R^{6} B^{2}}{I P_{\rm obs}} \sin^2\theta.
     \label{Eq:Pdot}
\end{equation}

For J2211+1136, adopting $R = 3210$ km, $I = (2/5) M R^2 = 1.04 \times 10^{50}$ g cm$^{2}$, and $B = 15$ MG, we obtain $\dot{P} \approx 3.3 \times 10^{-17} \sin^2\theta$ s s$^{-1}$ (see also \citealp{Williams_2022}). This spindown rate is too low to be detected, e.g., $2$ orders of magnitude lower than the one of the pulsating WD G 117-B15A, $\dot{P} \approx 5.12 \times 10^{-15}$ s s$^{-1}$ \citep{2021ApJ...906....7K}.

\subsection{Rotational evolution of J1901+1458} \label{subsec:ZTF}

J1901+1458 has a period of rotation $P_{\rm obs} = 416.2$ s a mass $M =$ $1.327$--$1.365\, M_\odot$, a stellar radius $R=2140^{+160}_{-230}$ km, and a surface (dipole) magnetic field in the range $B = 600$--$900$ MG. The cooling age is $t_{\rm cool} = 10$--$100$ Myr \citep{2021Natur.595...39C}. For simplicity, we consider $M = 1.35 \, M_\odot$ with $R=2140$ km in the analysis of this source. Furthermore, we follow the analogous procedure described above for J2211+1136. We analyze the evolution of the WD rotation period for different values of the initial rotation period, a disk mass of $M_{d} = 0.34 M_{\odot}$ consistent with the range of values obtained in Sec. \ref{sec:6}, and a value of $\dot{M}$ consistent with Eq.~(\ref{eq:Mdot}). We use a dipole magnetic field strength of $B = 800$ MG, inferred in \citet{2021Natur.595...39C} from the analysis of the position of the H$\rm \alpha$, H$\rm \beta$, and H$\rm \gamma$ spectral lines.

\begin{figure}
\includegraphics[width=\hsize,clip]{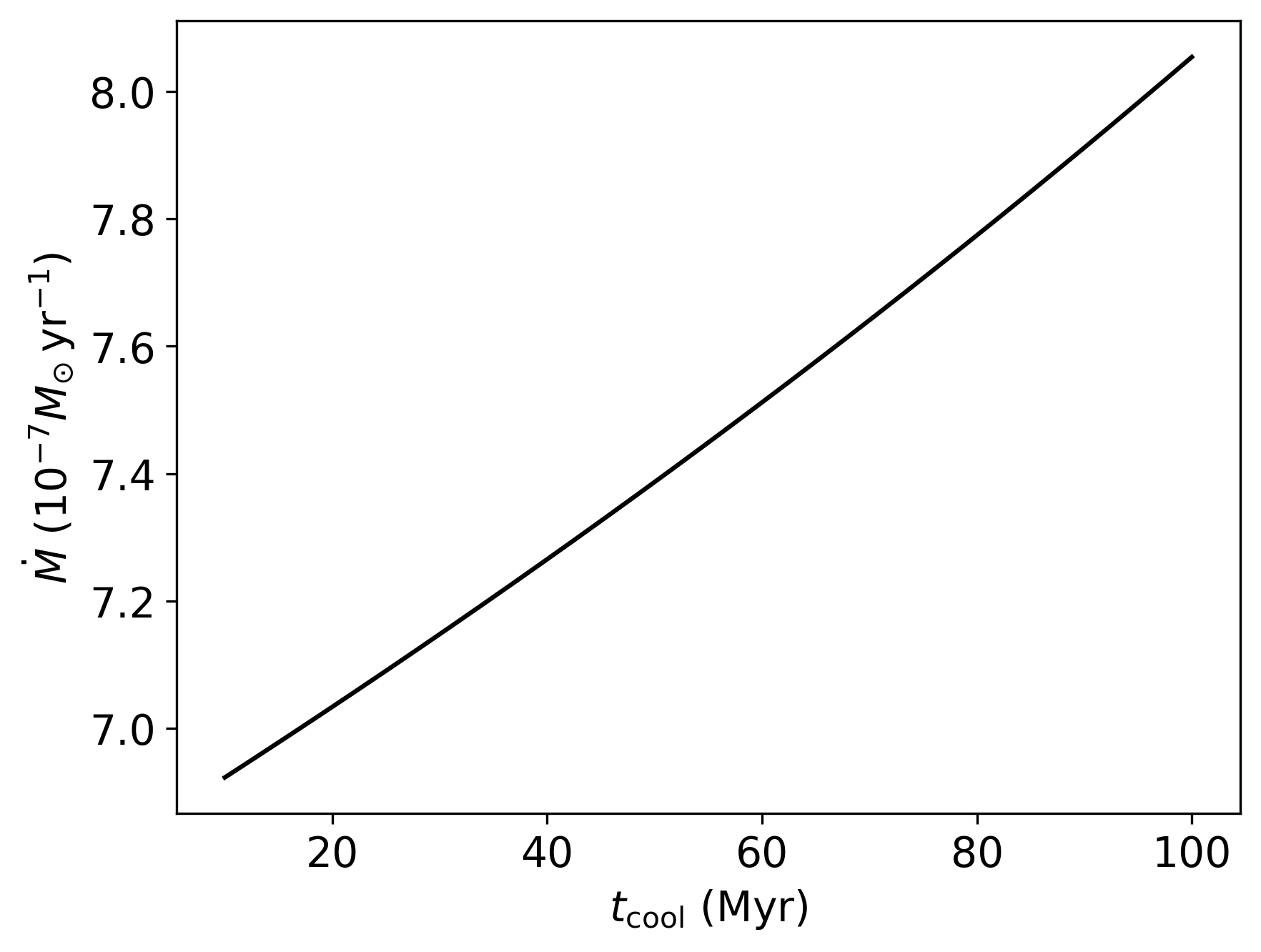}
\caption{Accretion rate as a function of cooling age calculated from Eq. (\ref{eq:Mdot}) for J1901+1458. The WD parameters are $M = 1.35 M_{\odot}$, $R = 2.14 \times 10^{8}$ cm, $P_{0} = 3.14$ s, $P_{\rm obs} = 416.2$ s. The magnetic field strength is $B = 800$ MG and  $\theta = 90^{\circ}$.}\label{fig:ZTF_ACC_time}
\end{figure}

\begin{figure*}
\includegraphics[width=\hsize,clip]{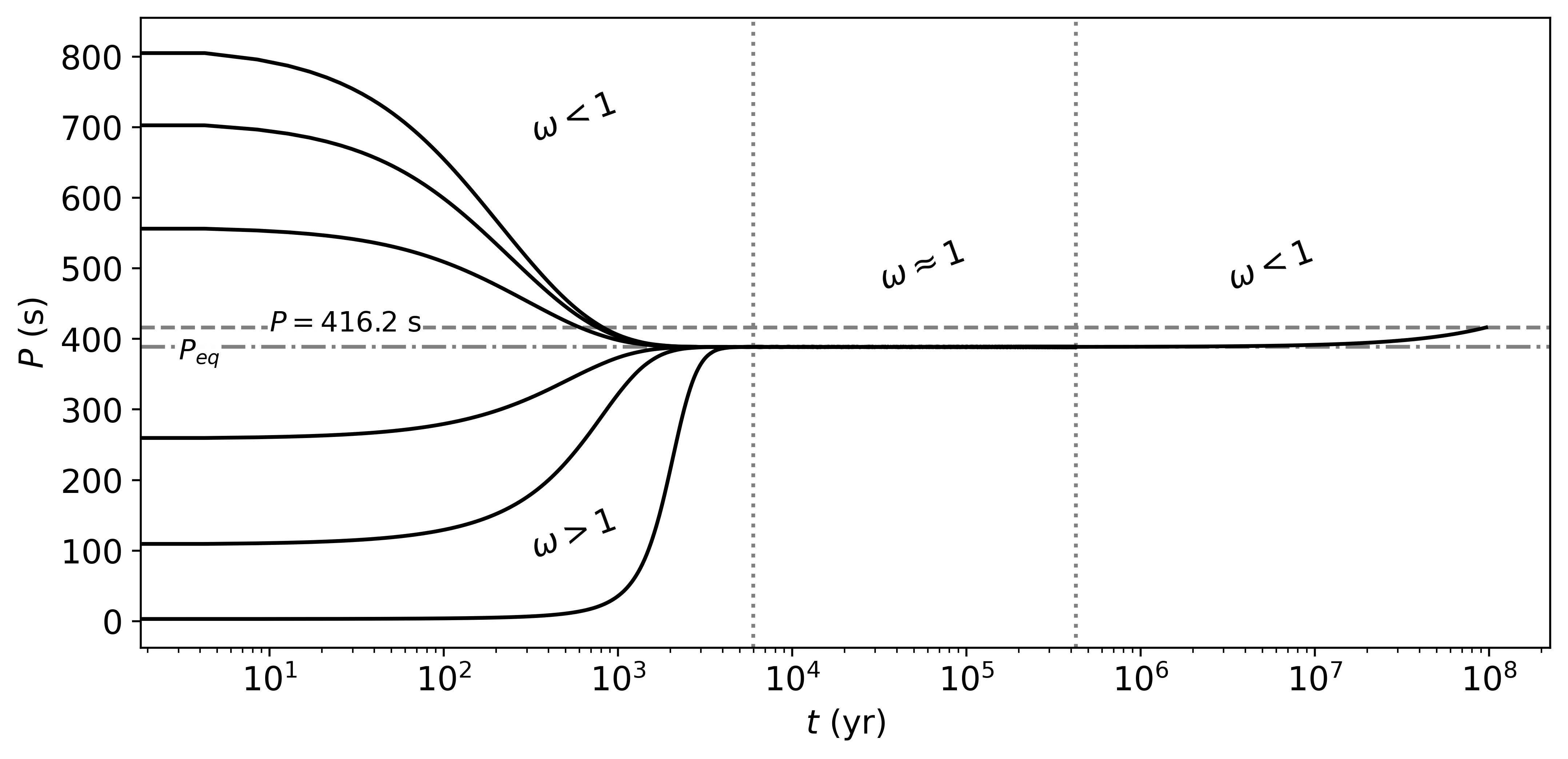}
\caption{Evolution of spin period of the J1901+1458 for $\dot{M} = 8.0 \times 10^{-7} M_{\odot}$ yr$^{-1}$ (with $B = 800$ MG) and for different values of initial rotational periods, $P_{0} = (3.14, \, 108.7, \, 258.7, \, 558.7, \, 708.7, \, 814.2)$ s. The dotted lines divide the evolution into three stages according to the value of $\omega$. In the first stage, WD can start with $\omega > 1$ or $\omega < 1$ depending whether the initial period is below or above the equilibrium period, $P_{\rm eq} = 388.6$ s (Eq. \ref{eq:Omeeq}), respectively. For both values, the involved torques are dipole radiation torque, $T_{\rm mag}$, and accretion torque, $T_{\rm acc}$, in the propeller phase. However, $T_{\rm acc}$ is the dominant torque in this phase. For $\omega \approx 1$, $T_{\rm mag} \approx T_{\rm acc}$ and the system goes through the equilibrium period, $P_{\rm eq}$. For $\omega < 1$, the involved torque is $T_{\rm dip}$. $T_{\rm acc}$ no longer acts on the star, as in the third stage, the disk has already exhausted. The upper dashed line indicates the current rotation period of the WD.}\label{fig:ZTF_twovalue}
\end{figure*}

Using Eq.~(\ref{eq:Mdot}), we obtain the accretion rate as a function of $t_{\rm cool}$. Considering $\theta = 90^{\circ}$, we obtain $\dot{M}\approx (6.92$--$8.05) \times 10^{-7} M_{\odot}$ yr$^{-1}$ (See Fig. \ref{fig:ZTF_ACC_time}). For an aligned rotator, $\dot{M}\approx (6.87$--$7.39) \times 10^{-7} M_{\odot}$~yr$^{-1}$. Figure \ref{fig:ZTF_twovalue} shows the rotation period evolution for for $\dot{M} = 8.0 \times 10^{-7} M_{\odot}$~yr$^{-1}$, and initial values of the rotation period $P_{0} = (3.14, \, 108.7, \, 258.7, \, 558.7, \, 708.7, \, 814.2)$~s. The curves approach the equilibrium period, $P_{\rm eq}$, in a timescale of the same order of magnitude. Thus, also for this source, we see that the final evolution time is not affected by the choice of the initial period.

Without loss of generality, we now describe the phases of the spin evolution in the case $P_{0} = 3.14$~s. The system first evolves through the propeller regime $\omega > 1$, in which the WD spins down to a period of $388.6$~s in $\Delta t_1\approx 5.98$~kyr (marked by the first dotted line in Fig. \ref{fig:ZTF_twovalue}). During this time, the amount of the disk mass ejected was $M_{\rm loss,1} = \dot{M}\Delta t_1 \approx 4.8 \times 10^{-3} M_{\odot}$.

After the rotation period reaches $\omega \approx 1$, the second stage starts. Thus, J1901+1458 evolves through spindown and spinup stages around the equilibrium period value, until the disk mass ends. We estimate that this phase lasts $\Delta t_2\approx 0.42$ Myr. Up to this point, the post-merger evolution time adding the two stages is $\Delta t_1+\Delta t_2\approx 0.425$ Myr, which is marked by the second dotted line in Fig. \ref{fig:ZTF_twovalue}. In this phase II, we have $M_{\rm acc,2} = M_{\rm loss,2} = (1/2)\dot{M}\Delta t_2 \approx 0.168 M_\odot$. Thus, in this source, nearly the entire disk mass is consumed in phase II, so we have a total ejected mass $M_{\rm loss} = M_{\rm loss,1} + M_{\rm loss,2} \approx 0.17 M_\odot$ and a total accreted mass $M_{\rm acc} = M_{\rm acc,2} \approx 0.17 M_\odot$.

In the subsequent evolution, the only torque acting onto the WD is $T_{\rm dip}$, so the WD enters the phase characterized by $\omega <1$. The WD spins down from a period of $388.6$~s to the observed period $P_{\rm obs}=416.2$~s. This occurs in $\approx 96.2$~Myr. Figure \ref{fig:ZTF_twovalue} shows this last stage for the spin evolution.

The spindown rate in the current phase for J1901+1458 can be estimated from Eq. (\ref{Eq:Pdot}). Adopting $R = 2140$ km, $I = (2/5) M R^2 = 5.0 \times 10^{49}$ g cm$^{2}$, and $B = 800$ MG, we obtain $\dot{P} \approx 2.9 \times 10^{-15} \sin^2\theta \,$s$\,$s$^{-1}$, which is consistent with the upper limit value of $\dot{P} < 10^{-11}$ s s$^{-1}$ presented in \citet{Caiazzo2021}. In this case, the resulting spindown rate is similar to the spindown rate of G 117-B15A. Therefore, although experimentally challenging, timing analyses as the one done by \citet{2021ApJ...906....7K} for G 117-B15A could lead in the future to the final proof of the DWD merger scenario for J1901+1458 presented in this work.

\section{The DWD merger progenitors}\label{sec:6}

We now turn to estimate the masses of the components ($M_1$ and $M_2$) of the DWD progenitor of the above two systems. For this task, we have to estimate the binary total mass, $M_{\rm tot}$, and mass-ratio, $q$, so that 
\begin{equation}\label{eq:M1M2}
    M_1 = \frac{1}{1+q}M_{\rm tot}, \qquad M_2 = \frac{q}{1+q}M_{\rm tot}.
\end{equation}

First, we use the fact that DWDs eject a tiny amount of mass during merger \citep{2009A&A...500.1193L,2014MNRAS.438...14D}, which allows us to assume baryon mass conservation with an error of at most one part in a thousand. Then, for a given disk mass, $M_d$, and accretion rate, $\dot{M}$, we have here obtained in the post-merger evolution the total accreted mass by the central WD remnant, $M_{\rm acc}$, and the total mass loss due to action of the propeller, $M_{\rm loss}$. Clearly, $M_d = M_{\rm acc} + M_{\rm loss}$. Therefore, we can write the total binary mass as
\begin{equation}\label{eq:Mtot2}
    M_{\rm tot} = M + M_{\rm loss} + M_{\rm ej} \approx M + M_{\rm loss},
\end{equation}
where we recall that $M$ is the current measured mass of the WD, and $M_{\rm loss}$ can be written as
\begin{equation}\label{eq:Mloss}
    M_{\rm loss} = \frac{M_{d} \pm M_{\rm cons}}{2},
\end{equation}
where the $\pm$ sign is used when the phase I starts with propeller ($\omega_0 >1$) or accretion ($\omega_0 <1$). The quantity $M_{\rm cons}$ is the disk mass consumed in the phase I. We now can find a equation for $M_{d}$ as a function of the mass ratio, $q$. To do this, we substitute Eq. (\ref{eq:Mtot2}) into Eq. (\ref{eq:md}) and solve it for $M_{d}$, which leads to
\begin{equation}\label{eq:Md_q}
    M_{d} = \frac{Q}{2-Q} \left ( 2M \pm M_{\rm cons} \right ),
\end{equation}
where $Q = -0.1185 + 0.9763q -0.6559q^{2}$. Furthermore, we can use Eq. (\ref{eq:Md_q}) together with Eq. (\ref{eq:Mloss}) to express $M_{\rm loss}$ in terms of $q$
\begin{equation}\label{eq:Mloss_q}
    M_{\rm loss} = \frac{Q}{2-Q} \left ( M \pm \frac{M_{\rm cons}}{Q} \right ).
\end{equation}

Therefore, with the equations obtained above, we can calculate $M_{\rm tot}$ and the masses of the components of the DWD progenitor with the following general procedure. Having chosen the initial rotation period and the accretion rate, we calculate the disk mass consumed in the phase I, $M_{\rm cons}$. Thus, the disk mass function $M_d = M_d(M_{\rm cons},q)$, given by Eq. (\ref{eq:Md_q}), and the mass loss function $M_{\rm loss} = M_{\rm loss}(M_{\rm cons},q)$, given by Eq. (\ref{eq:Mloss_q}), become a function of $q$. These functions are  concave-down parabolas with maximums at $q^*$, so the disk mass and mass loss increase with $q$ up to the maximum value $M^{\rm max}_d = M_d(q^*)$ and $M^{\rm max}_{\rm loss} = M_{\rm loss}(q^*)$, to then decrease up to the values $M_d(q=1)$ and $M_{\rm loss}(q=1)$, respectively. Therefore, if $M_d \leq M_d(q=1)$ or $M_{\rm loss} \leq M_{\rm loss}(q=1)$, there is a unique solution for $q$. If $M_d > M_d(q=1)$ or $M_{\rm loss} > M_{\rm loss}(q=1)$, there are two solutions for $q$.

We shall seek for solutions with a disk mass that satisfies $M_d \gtrsim M^{\rm min}_d$, where $M^{\rm min}_d = 0.1 M_\odot$ is approximately the minimum disk mass obtained in the numerical simulations of \citet{2014MNRAS.438...14D}, taking into account that $M \approx 1.3 M_\odot$ in the two analyzed systems.

We recall that in view of the nonzero disk mass left by DWD mergers, we have already discarded solutions in which the system evolves only through the phase III, i.e., only under the action of the magnetic dipole torque. Therefore, we are left with three possible cases: (i) evolution with phases I+III, (ii) with phases II+III, and the most general case (iii) with phases I+II+III. We now analyze each case.

\subsection{Case (i): evolution with phases I+III}

In this case, the system does not evolve through the phase II, so the entire disk mass is consumed in phase I, i.e., $M_{\rm cons} = M_{d}$. From Eq. (\ref{eq:Mloss}), we have that when phase I is a propeller ($\omega_0 >1$), $M_{\rm loss} = M_{d}$, while if it is an accretor ($\omega_0 <1$), $M_{\rm loss} = 0$. It can be shown that under these conditions, to satisfy the boundary condition of approximate equality of the rotational and the cooling age, either the disk mass must be $M_{d} \ll M^{\rm min}_d = 0.1 M_\odot$, or the accretion rate has very large values $\gtrsim 10^{-5} M_\odot$ yr$^{-1}$. Therefore, we do not consider this case as astrophysically viable.

\subsection{Case (ii): evolution with phases II+III}

In this case, the initial angular velocity satisfies $\Omega_{0} = \Omega_{\rm eq}$, so there is no phase I and the disk mass is divided in equal parts into accretion and propeller. We have $M_{\rm cons} = 0$, so Eq. (\ref{eq:Mloss}) leads to $M_{\rm loss} = M_d/2$.

\subsection{Case (iii): evolution with phases I+II+III}

In this general case, the system evolves through the three phases as described in Sec. \ref{sec:4}. The angular velocity at the beginning of the phase III, say $\Omega_{\rm dip}$, is approximately given by the equilibrium value, i.e. $\Omega_{\rm dip} \approx \Omega_{\rm eq}$. With this constraint, we can estimate $M_{\rm cons}$ by
\begin{equation}\label{eq:Mcons}
    M_{\rm cons} \approx  I \left ( \frac{\Omega_{\rm eq}^2}{GM} \right )^{2/3} \ln\left ( \frac{\Omega_0 - \Omega_{\rm eq}}{\Omega_{\rm dip} - \Omega_{\rm eq}} \right ),
\end{equation}
which is obtained from Eq. (\ref{eq:omega1}), solving it for $t$, making $t = \Delta t_{1} = M_{\rm cons}/ \dot{M}$
and $\Omega = \Omega_{\rm dip}$. Therefore, $M_{\rm cons}$ lies in the range $0 < M_{\rm cons} < M_{d}$.

\subsection{Specific examples}

\begin{figure}
    \centering
    \includegraphics[width=\hsize,clip]{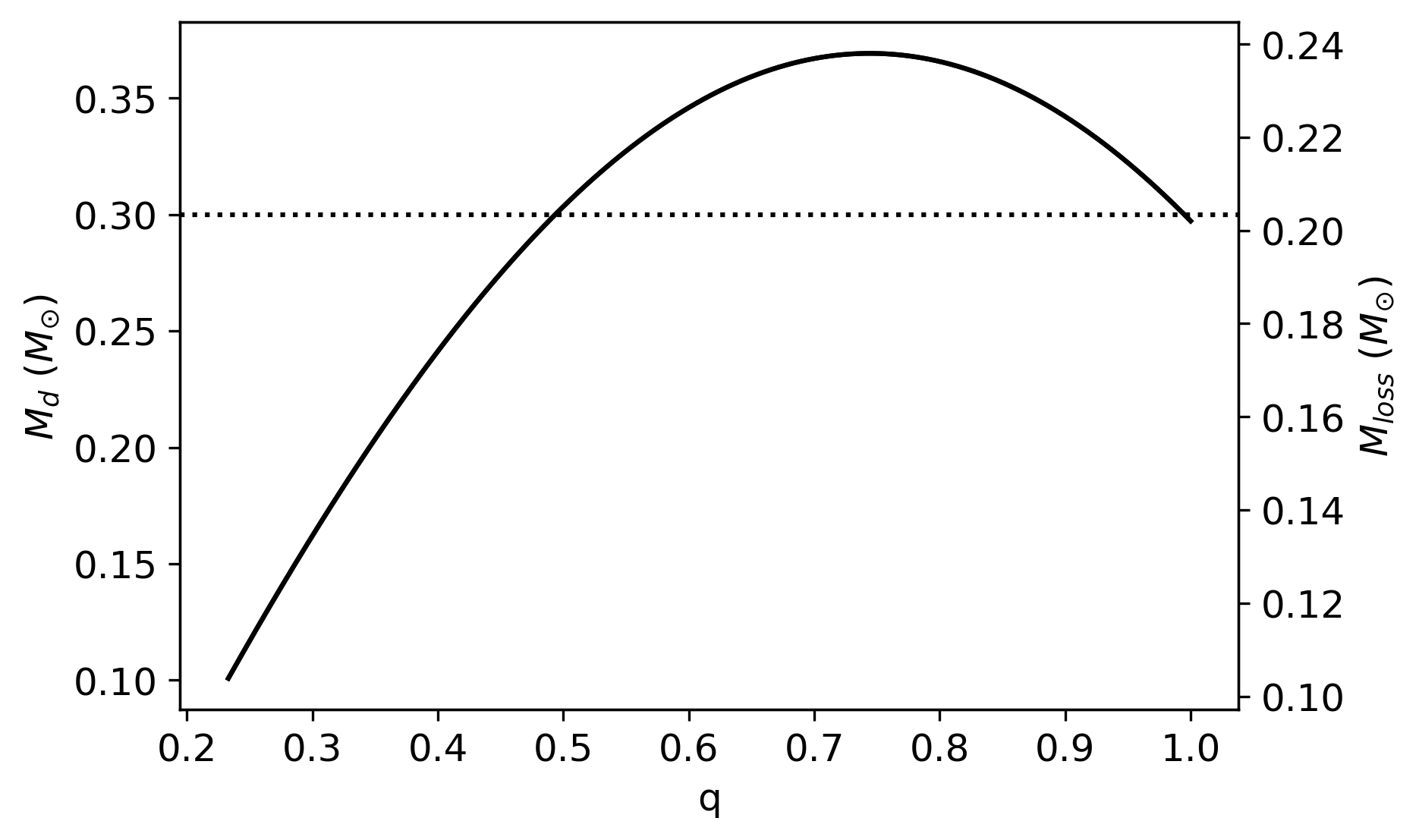}
    \caption{Disk mass, $M_d$, and mass loss, $M_{\rm loss}$, as a function of the binary mass-ratio, $q$. The solid curves represent the solutions for J2211+1136 ($M=1.27 M_\odot$; $P_{0} = 3.14$ s; $\dot{M} = 2.62 \times 10^{-7} M_{\odot}$ yr$^{-1}$), while the dashed curves show the corresponding used value in the simulation.}
    \label{fig:MdJ22}
\end{figure}

\begin{figure}
    \centering
    \includegraphics[width=\hsize,clip]{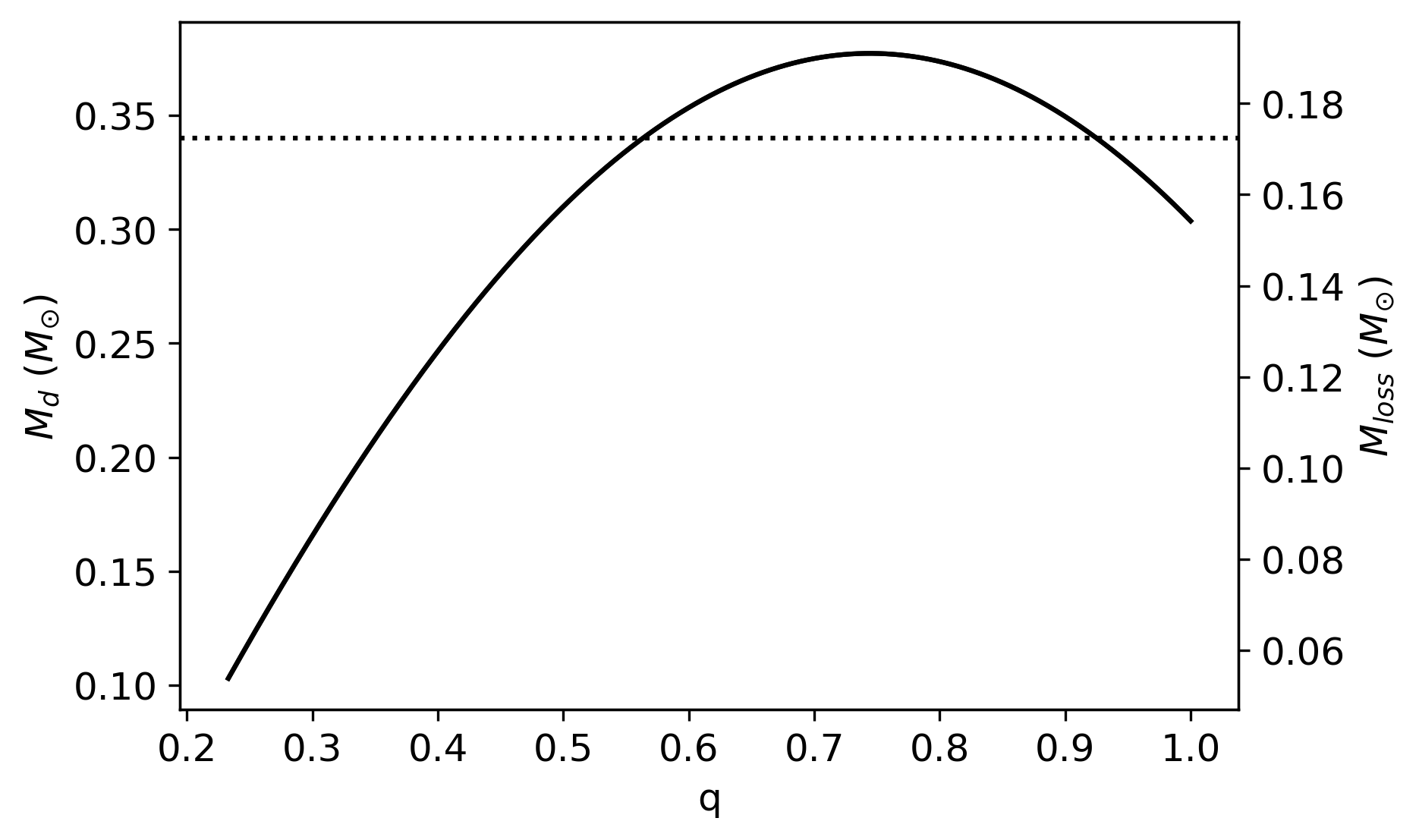}
    \caption{Disk mass, $M_d$, and mass loss, $M_{\rm loss}$, as a function of the binary mass-ratio, $q$. The solid curves represent the solutions for J1901+1458 ($M=1.35 M_\odot$; $P_{0} = 3.14$ s; $\dot{M} = 8.0 \times 10^{-7} M_{\odot}$ yr$^{-1}$), while the dashed curves show the corresponding used value in the simulation}.
    \label{fig:MdJ19}
\end{figure}

To exemplify the approach above, we consider the cases of J2211+1136 and J1901+1458 in Sec. \ref{sec:5}, i.e., in the general evolution (iii) in which the system evolves through phases I+II+III. Figure \ref{fig:MdJ22} shows the functions $M_d(M_{\rm cons},q)$ and $M_{\rm loss}(M_{\rm cons},q)$ for J2211+1136, given $M_{\rm cons} \approx 0.1 M_\odot$ ($P_{0} = 3.14$ s and $\dot{M} = 2.62 \times 10^{-7} M_{\odot}$ yr$^{-1}$). Figure \ref{fig:MdJ19} is analogous to Fig. \ref{fig:MdJ22} but for J1901+1458 with $M_{\rm cons} \approx 4.8 \times 10^{-3} M_\odot$ ($P_{0} = 3.14$ s and $\dot{M} = 8.0 \times 10^{-7} M_{\odot}$ yr$^{-1}$). Thus, for J2211+1136 and J1901+1458, we obtain $M^{\rm max}_d \approx 0.36 M_\odot$, $M_d(q=1) \approx 0.297 M_\odot$, and $M^{\rm max}_d\approx 0.37 M_\odot$, $M_d(q=1) \approx 0.304 M_\odot$, respectively.  From the $M^{\rm min}_d$ and $M^{\rm max}_d$ values obtained above, we note that our choice for the disk mass of J2211+1136, $M_d = 0.30 M_\odot$, and of J1901+1458, $M_d= 0.34 M_\odot$, are within the range of physically plausible values. The horizontal dashed curves in Figs. \ref{fig:MdJ22} and \ref{fig:MdJ19} represent these values and their intersection with the curve of the source gives the mass-ratio $q$ to these cases.


\begin{figure}
    \centering
    \includegraphics[width=\hsize,clip]{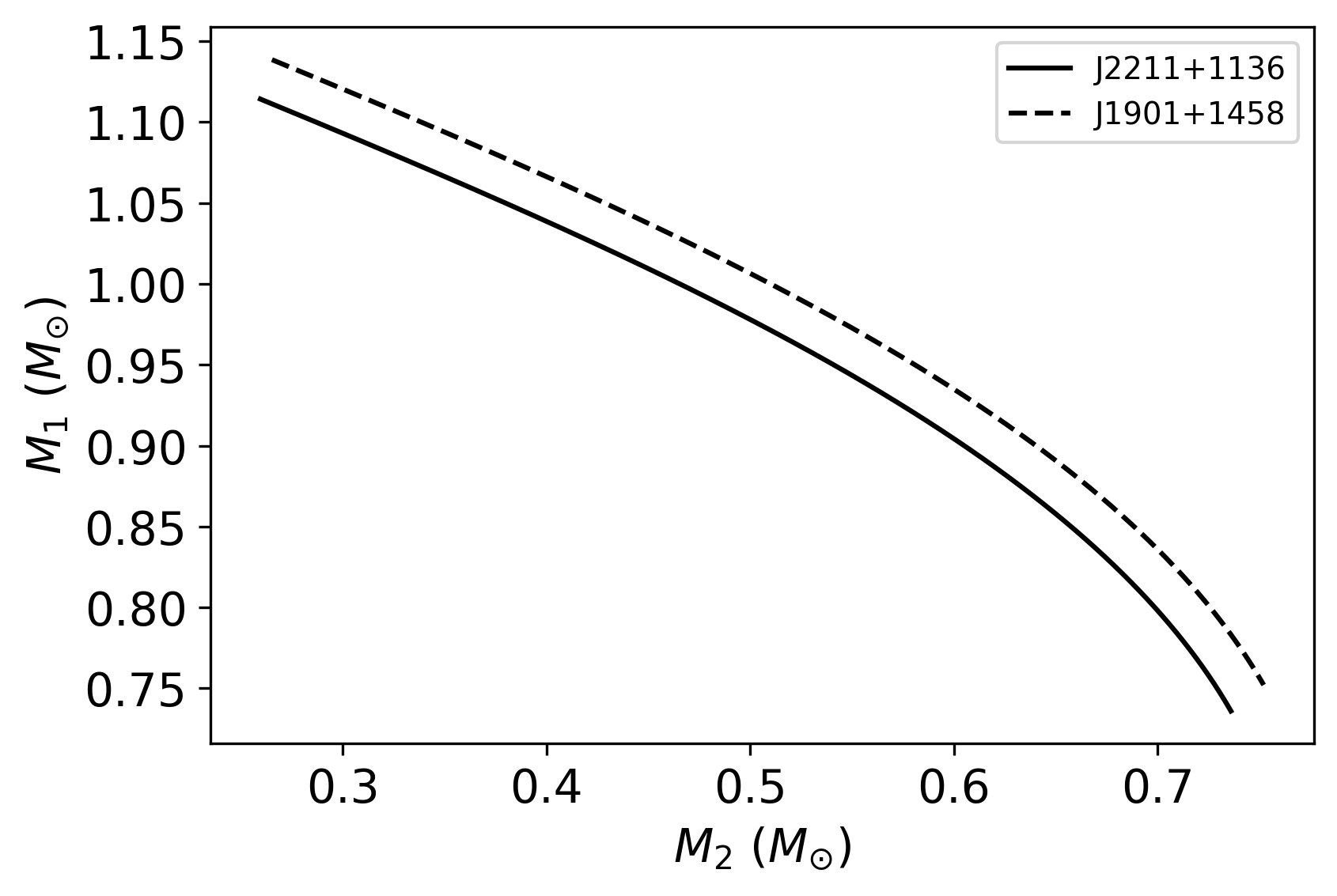}
    \caption{Predicted range of the primary and secondary mass. The solid curves represent the solutions for J2211+1136 (post-merger WD mass of $M=1.27 M_\odot$, $P_{0} = 3.14$ s and $\dot{M} = 2.62 \times 10^{-7} M_{\odot}$ yr$^{-1}$) and the dashed curves for J1901+1458 (post-merger WD mass of $M=1.35 M_\odot$, $P_{0} = 3.14$ s and $\dot{M} = 8.0 \times 10^{-7} M_{\odot}$ yr$^{-1}$)}.
    \label{fig:M1vsM2}
\end{figure}

Moreover, Figs. \ref{fig:MdJ22} and \ref{fig:MdJ19} show that given a value of $q$, we can estimate the $M_d$ and $M_{\rm loss}$, so we can calculate $M_{\rm tot}$ from Eq. (\ref{eq:Mtot2}). With these values, we can obtain the primary and secondary mass via Eq. (\ref{eq:M1M2}). In Fig.$\,$\ref{fig:M1vsM2}, we show the $M_1$-$M_2$ plane of possible solutions of the DWD progenitor. Each pair $(M_2, M_1)$ in this figure corresponds to a value of $q$ and, consequently, to a value of $M_d$. Therefore, we can infer the values $M_1$ and $M_2$ for the simulated cases in Sec. \ref{sec:5} taking into account the assumed value of $M_d$.

Therefore, since for the two systems we have chosen a disk mass value $M_d > M_d(q=1)$, there are two possible values of mass-ratio, $q_a$ and $q_b$ (see Figs. \ref{fig:MdJ22} and \ref{fig:MdJ19}). For J2211+1136, we obtain $q_a = 0.495$ and $q_b = 0.993$, while for J1901+1458, $q_a = 0.565$ and $q_b = 0.924$. For J2211+1136, the mass-ratio $q_a$ gives $M_1 = 0.983 M_\odot$, $M_2 =0.487 M_\odot$, while the solution $q_b$ gives $M_1 = 0.737 M_\odot$, $M_2=0.732 M_\odot$. For J1901+1458, the mass-ratio $q_a$ gives $M_1=0.971 M_\odot$, $M_2=0.549 M_\odot$, and $q_b$ gives $M_1 = 0.790 M_\odot$, $M_2=0.729 M_\odot$. Table \ref{tab:parameters} summarizes the results of above analysis. In particular, it lists the parameters of the pre-merger DWD and the parameters of the post-merger system for the disk mass and accretion rate in the simulations of Sec. \ref{sec:5}.
\begin{table}[ht]
    \centering
    \caption{Parameters of the pre-merger (DWD) and post-merger (central remnant WD+disk) systems leading to the current observed parameters of J2211+1136 and J1901+1458. There are two possible solutions for the mass of the DWD components, $M_1$ and $M_2$, corresponding to the two values of the binary mass-ratio, $q_a$ and $q_b$, that solve the system constraints (see Fig. \ref{fig:MdJ22} and \ref{fig:MdJ19}). $\Delta t_{\rm obs}$ is the rotational age, i.e., the total time elapsed since the merger to the instant when the WD reaches the current measured rotation period. We recall that we constrain the system to have $\Delta t_{\rm obs}$ equal to the estimated WD cooling age. See the text for further details.}
    \begin{tabular}{c|c|c}
      Parameter & J2211+1136 & J1901+1458\\
      \hline
      \multicolumn{3}{c}{Pre-merger system}\\
      \hline
       $q_a$, $q_b$ & $0.49$, $0.99$ & $0.56$, $0.92$ \\
       $M_{\rm tot}\,(M_\odot)$ & $1.47$ & $1.52$ \\
       $M_1\,(M_\odot)$ & $0.98$, $0.74$ & $0.97$, $0.79$ \\
       $M_2\,(M_\odot)$ & $0.49$, $0.73$ & $0.55$, $0.73$ \\
       $M_{\rm ej}\,(10^{-3} M_\odot)$ & $4.96$, $0.51$ & $3.61$, $0.68$\\
       \hline 
       \multicolumn{3}{c}{Post-merger system}\\
       \hline
       $M\,(M_\odot)$ & $1.27^{a,b}$ & $1.35^c$ \\
       $R$ (km) & $3210$ & $2140^c$ \\
       $P$ (s) & $70.32^{a}$ & $416.20^c$\\
       $B$ ($10^6$ G) & $15^{a}$ & $800^c$\\ 
       $B_{\rm quad}$ (G) & unconstrained & unconstrained \\
       $M_d\,(M_\odot)$ & $0.30$ & $0.34$ \\
       $\dot{M}\,(10^{-7}\,M_\odot$ s$^{-1})$ & $2.62$ & $8.0$\\
       $M_{\rm loss}\,(M_\odot)$ & $0.20$ & $0.17$ \\
       $\Delta t_{\rm obs}$ (Gyr) & $ 2.66 $ & $ 0.096 $ \\
       \hline
    \end{tabular}
    \tablerefs{$^a$\citet{2021ApJ...923L...6K}, $^b$\citet{2021MNRAS.503.5397K}, $^c$\citet{2021Natur.595...39C}.}
    \label{tab:parameters}
\end{table}

Interestingly, the inferred parameters of the DWD progenitor of J1901+1458 and J2211+1136 (see Table \ref{tab:parameters}) are consistent with the parameters of known systems, e.g., NLTT 12758, a $0.83+0.69 M_\odot$ DWD \citep{2017MNRAS.466.1127K}. This result further supports the link that we have here provided between DWD mergers and the formation of HFMWDs.

\section{Discussion and Conclusions}\label{sec:7}

In this article, we have investigated the possibility that the HFMWDs, J2211+1136, and J1901+1458, are DWD merger products. Based on numerical simulations of DWD mergers, we have modeled the post-merger system as a central WD surrounded by a disk from which there is a mass inflow towards the WD remnant. We have calculated the post-merger rotational evolution of the WD and inferred the system parameters for which the rotational age agrees with the WD cooling age.

We have shown that the post-merger configuration evolves through three different phases depending on whether accretion, mass ejection (propeller), or magnetic braking dominate the torque onto the WD. We have shown that the WD spends most of its lifetime in the third phase, in which only magnetic braking torques the WD (see Figs. \ref{fig:SDSS_26e8} and \ref{fig:ZTF_twovalue}). We have used the measured mass, magnetic field strength, and cooling age to infer the mass accretion rate and the disk mass.

The results of this article are the first attempt to establish a direct link between observed HFMWDs and their DWD merger progenitor. We conclude that the observed parameters of J2211+1136 and J1901+1458 are consistent with a DWD merger origin, and we have obtained the mass of the pre-merger DWD primary and secondary binary components (see Table \ref{tab:parameters}). Interestingly, the derived parameters of the merging DWDs are in line with those of known DWDs (like NLTT 12758; see \citealp{2017MNRAS.466.1127K}), which further supports the connection between HFMWDs and DWD mergers.

If HFMWDs like J2211+1136 and J1901+1458 are DWD merger products, the newborn merged remnant, besides being massive and highly magnetic, might be fast rotating in its early post-merger life (e.g., in its first $1$--$100$ kyr, see Figs. \ref{fig:SDSS_26e8} and \ref{fig:ZTF_twovalue}). WDs with such extreme properties might power a variety of transient and persistent electromagnetic phenomena in astrophysical sources. For instance, \citet{2021ApJ...906...53S} discusses evolutionary models of DWD merger remnants and how they might experience a $\sim 10$ kyr luminous giant phase during their final approach to the single massive WD or a neutron star fate.

The DWD merger and its early activity can also lead to low-energy gamma-ray bursts (GRBs). Phenomena in the merged magnetosphere can power the prompt gamma-ray emission, the cooling of the expanding ($\sim 10^{-3} M_\odot$) ejecta can power an infrared/optical transient days to week post-merger, and synchrotron emission of the ejecta and the WD pulsar-like emission can lead to an extended (years) X-ray, optical and radio emission \citep{2022IJMPD..3130013R, 2019JCAP...03..044R}.  

Massive, fast rotating HFMWD with pulsar-like activity might show up as magnetars \citep[see, e.g.,][and references therein]{2012PASJ...64...56M, 2013ApJ...772L..24R, 2014PASJ...66...14C, 2016JCAP...05..007M, 2017MNRAS.465.4434C, 2017A&A...599A..87C, 2018ApJ...857..134B, 2020ApJ...895...26B}. Other high-energy phenomena involving DWD mergers and HFMWD products are the emission of high-energy neutrinos \citep[see, e.g.,][]{2016ApJ...832...20X} and the particle acceleration leading to very-high ($\gtrsim 10^{15}$ eV)  and ultrahigh-energy ($\gtrsim 10^{18}$ eV) cosmic rays. In addition, space-based detectors of gravitational waves (GWs) like the Laser Interferometer Space Antenna (LISA) expect to detect the GW radiation driving the dynamics of compact, detached DWDs \citep[see, e.g.,][]{Stroeer2006Sep, Korol2022Apr}. Electromagnetic radiation phenomena might affect the evolution of merging DWDs detectable through the deviations from the case when pure GW radiation drives the orbital dynamics \citep[see][and references therein]{2022arXiv220800863C}, and the fast rotation and high magnetic fields might also lead to GW radiation from HFMWD pulsars \citep[see, e.g.,][]{2020MNRAS.492.5949S, 2020MNRAS.498.4426S}.


\vspace{0.4cm}
{\it{Acknowledgments:} We thank the Referee for the thoughtful comments and suggestions that helped us to improve the presentation of our results. M.F.S. thanks CAPES-PrInt (88887.351889/2019-00) for the financial support. J.G.C. is likewise grateful for the support of CNPq (311758/2021--5) and FAPESP (2021/01089--1), and financial support by ``Fen\^omenos Extremos do Universo" of Funda\c{c}\~ao Arauc\'aria. J.C.N.A. thanks CNPq (308367/2019-7) for partial financial support.}


\bibliography{sample631,references}
\bibliographystyle{aasjournal}



\end{document}